\documentstyle[proc,epsf]{rs}
\input epsf
\makeatletter

\def\compoundrel#1\over#2{\mathpalette\compoundreL{{#1}\over{#2}}}
\def\compoundreL#1#2{\compoundREL#1#2}
\def\compoundREL#1#2\over#3{\mathrel
  {\vcenter{\hbox{$\m@th\buildrel{#1#2}\over{#1#3}$}}}}
\makeatother

\begin{document}


\title[Continuous error correction]{Continuous error correction}

\author[J.P. Paz and W.H. Zurek]
{Juan Pablo Paz $^{1,3}$ and Wojciech Hubert Zurek $^{2,3}$}

\affiliation{$^{1}$ Departamento de F{\i}sica J.J. Giambiagi, 
FCEN, UBA, Pabell\'on 1, Ciudad Universitaria, 1428 Buenos Aires, Argentina\\
$^{2}$ Theoretical Astrophysics, MSB288, Los Alamos National Laboratory,
Los Alamos, NM 87545, USA\\
$^{3}$ Institute for Theoretical Physics, University of California, 
Santa Barbara, CA 93106-4030}

\maketitle

\begin{abstract}
{We propose a new approach to study the evolution of a quantum
state that is encoded in a system which is continuously subject to 
the operations required to implement a quantum error correcting code. 
In the limit of {\em continuous error correction} we introduce a 
Markovian master equation that includes the effects of: a) Hamiltonian 
evolution, b) errors caused by the interaction with an environment 
and c) error--correcting operations. 
The master equation is formally presented for all stabilizer
codes and its solution is analyzed for the simplest such code.}
\end{abstract}

\date{\today}

%
%

\section{Introduction}

In the last two years there has been enormous progress in the 
development of quantum error correction codes. The basic goal of 
these techniques is to protect quantum information against the 
degrading effect of uncontrolled interactions with external agents. 
To achieve this goal, being able to partially 
restore the unknown quantum state, one should perform a 
sequence of operations on the system used to represent the information. 
This active process, a 
quantum error correcting code, involves the following basic ingredients:
First, one should encode the information (initially contained in the quantum
state of $k$ qubits) into an entangled quantum state of $n$ carriers 
($n>k$). After the encoding, the original state 
is protected from certain errors provided one acts on the 
system applying a so--called recovery operator. The simplest (but not the 
only) way to describe this operation is to decompose it in the following 
steps: one should first perform a measurement 
on $n-k$ qubits, whose result reveals 
a syndrome. This measurement is followed by a unitary operation on 
the remaining $k$ qubits, whose nature depends upon the syndrome. Finally,
the operation is completed by reseting the $n-k$ measured qubits to 
a reference (ground) state. 
The short history of quantum error correcting techniques is quite 
remarkable: after the first pioneering papers by P. Shor and A. Steane
(Shor, 1995; Steane 1996$a,b$), several papers helped in clarifying
the nature of quantum codes (Laflamme \et , 1996; Bennett \et , 1996). 
Very soon, the mathematical structure behind these simple physical ideas
was revealed (Knill \& Laflamme, 1997) 
and the, rather general, theory of stabilizer codes was fully developed
(Gottesman, 1996$a$, 1997; Calderbank \et, 1996). More recently, it was 
shown that the use of concatenated quantum error correcting codes 
(Knill \& Laflamme, 1996) 
together with fault tolerant techniques for quantum computation (Shor, 1997)
may allow arbitrarily long computation provided the accuracy per
operation is below certain computational threshold (Knill, Laflamme \&
Zurek, 1997; Preskill, 1997). 

In this paper we propose a new tool that could be used to study the 
evolution of an encoded quantum state. The motivation of our method
lies on the interesting analogy between the active mechanism of 
quantum error correction and the process of cooling a system 
pumping energy away from it by driving certain internal transitions. 
Thus, we propose to use a master equation (inspired in the quantum--optical 
master equation) to describe the limit in which a system formed by 
$n$--carrier qubits is continuously being affected by the active 
process of error 
correction while, at the same time, is suffering the continuous 
interaction with an environment and, eventually, is evolving under 
a specified Hamiltonian. 
The limit of continuous error correction is the one in which one 
assumes that all operation involved in the code (encoding, applying 
the recovery operator, refreshing the syndrome, etc) can be considered
to be instantaneous for all practical purposes. 

Our purpose here is twofold. On the one hand we will introduce the 
general master equation for a rather large class of error correcting 
codes (stabilizer codes). On the other hand, as we will show that
this master equation can be explicitly solved in some simple but 
relevant cases, we will discuss important features of the 
solution. We find it rather interesting that quantum error correcting
codes admit a description in terms of a continuous differential equation. 
But it is certainly most remarkable that this equation not only can be 
formally written in terms of some abstract operators but can also be 
explicitly solved in some interesting cases. 
The rest of the paper is organized as follows: 
In Section 2 we present a very brief introduction to the theory of 
stabilizer error correcting codes. The ideas discussed in this section 
enable us to introduce the notion of ``corrective quantum jumps". 
In Section 3, this notion is used to motivate a master equation 
describing the evolution of a quantum system which is continuously 
being affected by errors and is subject to a continuous sequence of 
corrective operations. We analyze and solve the 
master equation for the simplest example of a stabilizer code: the $n=3$ 
qubit code protecting $k=1$ qubit of information against phase errors 
affecting any one of the three carriers. In this case we present  
the explicit solution of the master equation computing 
the fidelity from the density matrix of the system. 
We discuss the various time scales that 
affect the evolution of the encoded state in the collective Bloch sphere.

\section{Stabilizer error correcting codes.}

We will consider a quantum error correcting code that protects $k$ qubits by 
encoding them into $n$ carriers. The code space ${\cal H}_k$ is a $2^k$ 
dimensional
subspace of the Hilbert space of the $n$ carriers. ${\cal H}_n$ is a 
tensor product 
of $n$ $2$--dimensional factors and has a natural basis whose elements
are product states of the individual carriers. 
This ``physical basis" can be formed with the common eigenstates
of the operators $\{Z_1,\ldots,Z_n\}$ ($X_i, Y_i$ and $Z_i$ denote the  
Pauli matrices for the $i$--th qubit). We will label states of this basis not 
by the eigenvalues of the corresponding operators (which are $\pm 1$) 
but by the 
eigenvalues of the projectors onto the $-1$ subspace (which are $0$ 
or $1$): Thus, 
the label $z_j=0$ ($z_j=1$) corresponds to a $+1$ ($-1$) eigenvalue of the 
operator $Z_j$. Furthermore, we order the $n$ carriers in 
such a way that the last $k$ qubits are the ones whose state we encode 
while the 
first $n-k$ are the ancillary carriers. Therefore, states of the physical 
basis are of the form $|s,z\rangle_P=|s\rangle_P\otimes|z\rangle_P$ 
(where the strings $s=(s_1,\ldots,s_{n-k})$, $z=(z_1,\ldots,z_k)$ 
store the corresponding eigenvalues and the subscript $P$ is used to identify
states of the physical basis). 

An error correcting code is a mapping from the physical product 
states $|0\rangle_P\otimes |\Psi\rangle_P$ onto the code 
space ${\cal H}_k$, which is 
formed by entangled states of the $n$ carriers. A rather general 
class of codes can 
be described in terms of their stabilizer group (see Gottesman, 1997). 
The stabilizer of the code is an Abelian group formed by all operators which 
are tensor 
products of Pauli matrices and have ${\cal H}_k$ as an eigenspace with 
eigenvalue $+1$. 
Every element of the stabilizer, which is a finite group with $2^{n-k}$ 
elements, 
can be obtained by appropriately multiplying $n-k$ generators, which will be
denoted as $M_1,\ldots, M_{n-k}$. The elements of the stabilizer are completely
degenerate in the code space ${\cal H}_k$ (since all states in ${\cal H}_k$ 
are eigenstates with 
eigenvalue $+1$ of all $M_j$). To define a basis in the code space we 
choose $k$ operators $L_1,\ldots,L_k$ which being tensor products of 
Pauli matrices 
commute with all elements of the stabilizer. 
These operators $L_{j'},\ j'=1,\ldots,k$ are the ``logical pointers" 
since they 
define the directions in ${\cal H}_k$ associated with the logical states  
$|0>_L,\ldots, |2^k-1>_L$ (logical pointers belong to the group of operators 
which commute with the stabilizer, known as the normalizer). 

The $n-k$ generators of the stabilizer together with the $k$ logical 
pointers are 
a Complete Set of Commuting Operators (CSCO) whose common eigenstates form 
a complete basis of the Hilbert space ${\cal H}_n$. Elements of 
this ``logical basis", labeled 
by $n$ quantum numbers, are denoted as $|m,l\rangle_L$, where the bit 
strings $m=(m_1,\ldots,m_{n-k})$, $l=(l_1,\ldots,l_k)$ identify 
the corresponding 
eigenvalues and the subscript $L$ refers to logical states.  
The CSCO formed by the generators of the stabilizer and the logical 
pointers defines 
a prescription for decomposing the original Hilbert space of 
the $n$--carriers into a 
tensor product of a $2^k$--dimensional logical space ${\cal A}$ and a 
$2^{n-k}$--dimensional syndrome 
space ${\cal S}$. In fact, elements of the logical 
basis (which are entangled 
states of the $n$-carriers) are tensor products of states belonging 
to ${\cal A}$ and 
${\cal S}$: 
$|m,l\rangle_L=|m\rangle_L\otimes|l\rangle_L$. Encoded states, 
which belong to $H_k$, 
are also product states of the form 
$|\Psi\rangle =|0\rangle_L \otimes\sum_l c_l |l\rangle_L$. 

The code protects quantum states against any error $E_a$ whose action on 
states of the
logical basis is to change the logical syndrome and, eventually, rotate 
the logical 
state in ${\cal A}$ in a syndrome dependent way:
\begin{equation}
E_a\ |m\rangle_L\otimes |l\rangle_L=
e^{i\phi_{ma}}\ |m+c_a\rangle_L\otimes U_a |l\rangle_L.
\end{equation}
Here $U_a$ is a unitary operator acting on the collective logical space $A$
and $\phi_{ma}$ is a phase that may depend on the syndrome 
and the error. The error 
$E_a$ changes the syndrome from $m$ to $m+c_a$ where $c_a$ is the bit string 
storing
the commutation pattern between the error and the generators of the
 stabilizer (the
$j$--th bit of this string is one if the error anti--commutes with $M_j$ and 
is zero otherwise). The reason for this is that when acting on a 
logical state, 
the error $E_a$ changes the eigenvalue of the operator $M_j$ 
only if $\{M_j,E_a\}=0$. The label $a$ used to identify errors is 
arbitrary and, 
for the case of non--degenerate codes (which is the only one we will 
consider here) 
it is always possible to label errors $E_a$ using simply the commutation 
pattern $c_a$ (i.e., we can choose $a=c_a$). 

To correct against the action of any of the errors $E_a$ (or against any 
linear superposition of them) one
can first detect the error by measuring the collective 
syndrome (i.e., measuring
the observables $M_j$, $j=1,\ldots,n-k$) and later recover from
the error by applying the corresponding operator $U^\dagger_a$.  
This detection--recovery process can be conveniently described as a 
quantum operation defined by the following mapping from the erroneous density 
matrix $\rho_{in}$ into the corrected one $\rho_{out}$:
\begin{equation}
\rho_{out}=\sum_{m=0}^N R_m\rho_{in} R^\dagger_m\label{recovermap}
\end{equation}
where the sum runs over all syndromes ($N=2^{n-k}-1$) and the recovery 
operator 
is 
\begin{equation}
R_m=|0\rangle_L\  _L\langle m|\otimes U^\dagger_m.\label{recoverop}
\end{equation}
By construction, these operators satisfy the identity 
$\sum_{m=0}^N R^\dagger_m R_m = I$.

As our description of  error detection--recovery process  is entirely 
formulated in the 
logical basis it does not involve any reference to encoding--decoding 
operations which
can be simply defined as a change of basis: The encoding operator $C$ 
is a unitary operator mapping the physical basis, formed by product 
states of the 
$n$ carriers, onto the logical basis, formed by entangled states. 
Accordingly, $C$ 
transforms the operators $Z_i$ (whose eigenvalues define states in 
the physical 
basis) into the operators $M_j$, $L_{j'}$ (that label states in the 
logical basis). 
Thus, the encoding operator $C$ is such that 
$Z_j=C^\dagger M_j C$, $j=1,\ldots,n-k$ and $Z_{n-k+j'}=C^\dagger L_{j'} C$, 
$j'=1,\ldots,k$. Taking this into account the action of the operator $R_m$ can 
be described, in the physical basis, as the following sequence of operations: 
i) decode the state, ii) measure the syndrome in the physical basis by 
measuring $Z_j$ in 
the first $(n-k)$ carriers, iii) If the result of the measurement is the 
string $s$, 
apply the syndrome dependent recovery operator $U^\dagger_s$ 
resetting the syndrome back to zero, iv) encode the resulting state. 

Finding a stabilizer code correcting a given set of errors is a 
rather hard task which involves designing generators having appropriate 
commutation 
patterns with the errors. Once the generators are found and
the logical pointers are chosen an encoding--decoding operator can be 
constructed 
(strategies for designing encoding--decoding circuits from the stabilizer are 
known; see Gottesman, 1997). The recovery operators depend on 
the encoding--decoding strategy and can be explicitly found from the 
encoding circuit 
by running errors through it. 

In the next Section we will illustrate our model for continuous error 
correction 
using the simplest stabilizer code: the one encoding $k=1$ qubit 
using $n=3$ carriers 
correcting against phase errors in any of the carriers. The basic errors 
are $E_1=Z_1$, $E_2=Z_2$ and $E_3=Z_3$. The stabilizer of the code can 
be chosen to be generated by: $M_1=X_1X_3$ and $M_2=X_2X_3$.
The commutation pattern associated with each error is $c_1=01$ (because 
the error $Z_1$ 
commutes with $M_1$ and anti--commutes 
with $M_2$), $c_2=10$, $c_3=11$ (note that we are
labeling errors with the commutation pattern). An encoding
circuit is shown in Figure 1. The operator $C$ has the following simple 
properties:
\begin{equation}
C^\dagger Z_1 C=X_1,\ \ C^\dagger Z_2 C=X_2,\ C^\dagger Z_3 C= X_1X_2Z_3.
\end{equation}
These properties entirely determine the action of the errors $Z_i$ in the 
logical basis. For example, the last identity implies that
$E_3|m\rangle_L|l\rangle_L=(-1)^l|m+c_3\rangle_L|l\rangle_L$. 
Thus, the error $E_3$ not only changes the syndrome but also 
modifies the logical state by adding a phase. This means that the recovery 
operator for this error is $U_3=Z$ 
Analogously, we can find how the other errors act on the logical 
basis showing that $U_1=U_2=I$.

\begin{figure}
\vspace {1.0cm}
\hspace {1.5cm}
\epsfxsize=8.6cm
\epsfbox{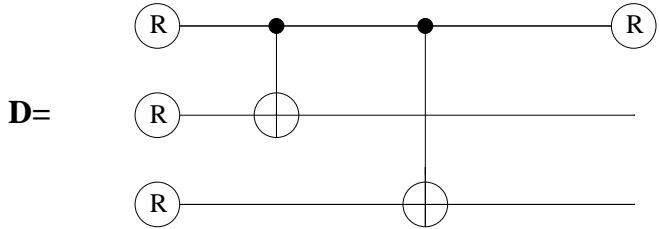}
\vspace {0.5cm}
\caption{Encoding circuit for the three qubit code protecting 
against phase errors in the physical basis. Qubits are
labeled from bottom to top. The third qubit is the one to be encoded. When 
run from left to right the circuit defines the decoding operator $D$. The 
encoding operator corresponds to the same circuit run backwards.}
\end{figure}

\section{Master equation with corrective quantum jumps.}

Quantum error correction is an active process: for the density matrix 
to evolve according to the rule (\ref{recovermap}), a sequence of physical 
operations must be carried out on the system (measuring the 
syndrome, refreshing it and applying the syndrome dependent recovery 
operator). 
Let us suppose that we are able to perform detection--recovery operations 
arbitrarily fast (i.e., the time required for these operations is 
negligible with 
respect to any other time scale 
in the system) and that the corrective protocol is
repeatedly applied many times. In the limit of continuous error correction 
we will consider that the time required to perform the corrective operations 
goes to zero while the number of corrective events goes to infinity. 
In this limit the 
evolution of the density matrix is generated by the continuous application 
of the mapping (\ref{recovermap}). 

In this paper we suggest that this rather idealized scenario, which 
might be useful 
to analyze the optimal protection achievable with a given code, 
could be described using a simple master equation. To motivate the 
master equation
one could notice that the mapping from the erroneous to the corrected 
density matrix (given by (\ref{recovermap})) is naturally described 
in the language 
of quantum jumps: In fact, the operators $R_m$ define ``corrective jumps" 
which reset 
the logical syndrome to $m=0$ while acting as $U^\dagger_m$ in the logical 
space ${\cal A}$. 
Thus, to describe the continuous limit we propose to use a Markovian 
master equation
(analogous to the quantum optical master equation, see for example 
Walls \& Milburn, 1994) having $N+1$ decay channels associated 
with the $R_m$--quantum jump operators:
\begin{equation}
\dot\rho=L_0\rho + \sum_m \gamma_m \left(R_m\rho R^\dagger_m - 
{1\over 2} R^\dagger_m R_m \rho - {1\over 2} \rho R^\dagger_m R_m\right).
\label{master1}
\end{equation}
Here, the constants $\gamma_m$ measure the strength of the corrective 
jumps and $L_0\rho$ includes all other sources of evolution (produced 
by a Hamiltonian $H_0$ or caused by the errors $E_a$). 
To write this non--corrective piece of the master equation 
we will consider a system with a Hamiltonian $H_0$ that is 
interacting with an environment. We assume that this interaction 
produces a term in 
the master equation which, in the Markovian limit, has the same form than
(\ref{master1}) with $N$ decay channels $E_a$:
\begin{equation}
L_0\rho=-i[H_0,\rho]+ \sum_{a=1}^N \gamma'_a \left(E_a\rho E^\dagger_a - 
{1\over 2} E^\dagger_a E_a \rho - {1\over 2} \rho E^\dagger_a E_a\right).
\label{L0}
\end{equation}

The complete master equation can be written in a rather compact form if we 
assume that all 
corrective jumps have equal strength ($\gamma_m=\gamma$) and that the same 
is 
true for the erroneous jumps ($\gamma'_a=\gamma'$). In this case, and if 
the erroneous jump operator are unitary, the complete master equation reads:
\begin{equation}
\dot\rho=-i[H_0,\rho] - (\gamma+N \gamma') \rho + 
\gamma \sum_{m=0}^N  R_m\rho R^\dagger_m 
+\ \gamma'\ \sum_{a=1}^N  E_a\rho E^\dagger_a.\label{mastereq}
\end{equation}

To solve this equation it is convenient to remember that, as the Hilbert
space ${\cal H}_n$ is a tensor product of a collective syndrome space 
$\cal{S}$ and a 
collective logical space $\cal{A}$, we can always write the density matrix 
$\rho$ as 
\begin{equation}
\rho=\sum_{m,m'=0}^N|m\rangle_L {}_L\langle m'| \otimes \rho_{mm'},
\end{equation}
where the operators $\rho_{mm'}$ act on the $2^k$--dimensional 
space $\cal{A}$. 
For $\rho$ to be hermitian and normalized, the following identities must be
satisfied: i) $\rho_{mm'}^\dagger=\rho_{m'm}$ and ii) 
$\sum_{m}Tr\left(\rho_{mm}\right)=1$. Using this notation one can 
rewrite both 
the corrective and the error--generating terms in (\ref{mastereq}) as
\begin{eqnarray}
\sum_{m=0}^N  R_m\rho R^\dagger_m &=& |0\rangle_L {}_L\langle 0|\otimes 
\sum_{m=0}^N U^\dagger_m \rho_{mm} U_m,\label{correctterm2}\\  
\sum_{a=1}^N  E_a\rho E^\dagger_a &=& \sum_{a=1}^N\sum_{mm'=0}^N 
|m+a\rangle_L {}_L\langle m'+a| \otimes 
e^{i\left(\phi_{ma}-\phi_{m'a}\right)} 
U_a\rho_{mm'} U^\dagger_{a},\label{errorterm2}
\end{eqnarray} 
where errors are labeled using the commutation pattern $c_a$ (i.e. we choose 
to take $a=c_a$). 
These equations show that neither the erroneous nor the corrective quantum 
jumps 
couple diagonal operators ($m=m'$) with off diagonal ones ($m\neq m'$). 
Using them, equations for diagonal operators $\rho_m=\rho_{mm}$ can be easily 
found. To explicitly write down the equations for $\rho_m$ we will assume
that the Hamiltonian $H_0$ does not connect states with different 
syndromes. For
this type of Hamiltonian (which are the ones required for a fault tolerant 
computation) we have $H_0=\sum_{m=0}^N|m\rangle\langle m|\otimes h_m$, 
where $h_m$
is a Hamiltonian acting in the logical subspace ${\cal A}$. Using this, 
the master equation which includes the effects of Hamiltonian evolution, 
errors and error correction, reads:
\begin{eqnarray}
\dot\rho_{m}&=&-i\left[h_m,\rho_m\right]
-(\gamma(1-\delta_{m0})+N\gamma')\rho_{m}+\nonumber\\
&+&\sum_{a=1}^N\left(\delta_{m0}\gamma U^\dagger_{a} \rho_{a} U_{a} + 
\gamma' U_a\rho_{m+a} U^\dagger_{a}\right).\label{masterdiag} 
\end{eqnarray}
As we can see from (\ref{masterdiag}), all operators tend to be damped by the 
corrective term except $\rho_0$. On the other hand, errors tend to induce 
transitions between different syndromes at a rate proportional to $\gamma'$. 
From these equations we can first compute the probability for detecting the 
$m$--th syndrome, which is nothing but the 
trace of $\rho_m$: 
defining $p_m=Tr(\rho_m)$ and taking the trace in (\ref{masterdiag}) we find:
\begin{equation}
\dot p_{m}=-(\gamma(1-\delta_{m0})+N\gamma') p_{m} +
\sum_{a=1}^N\left(\delta_{m0}\gamma p_{a} + \gamma' p_{m+a} \right).
\end{equation}
Solving this system, we obtain the probability for the correct syndrome
\begin{equation}
p_0 (t) = {\gamma+\gamma'\over{\gamma+(N+1)\gamma'}}+\left(p_0(0)-
{\gamma+\gamma'\over{\gamma+(N+1)\gamma'}}\right)
\exp(-\lambda t),\label{p0}
\end{equation}
where the decay rate is $\lambda=\gamma+(N+1)\gamma'$. This rather simple
formula, valid for all stabilizer codes in the limit of continuous 
correction, shows that the probability for detecting the correct 
syndrome tends to be of order unity if: i) the corrective coupling constant 
dominates over $\gamma'$ ($\gamma\gg\gamma'$ defines the strong 
correction limit) and
ii) times are large as compared to $t_c=1/\gamma$. 

To solve the master equation finding the operators $\rho_m$ we will be more 
specific restricting ourselves to consider codes protecting $k=1$ qubit. 
In this case the operator $\rho_m$ acts on a $2$--dimensional Hilbert space 
and can be written as a linear combination of Pauli matrices:
\begin{equation}
\rho_m={1\over 2}\left(p_m \hat 1+ \vec{r}_m \hat{\vec{\sigma}}\right).
\label{rhom}
\end{equation}
where $\hat{\vec{\sigma}}$ is the vector of Pauli matrices and 
the polarization vector $\vec r_m$ is the expectation value of  
$\hat{\vec{\sigma}}$ in the state $\rho_m$ (note that $\rho_m$ is 
not normalized 
since $p_m=Tr(\rho_m)$ is the probability for detecting the $m$--th syndrome). 
The master equation (\ref{masterdiag}) translates into a linear system of 
equations for the polarization vectors $\vec r_m$. Thus, the evolution of 
the quantum 
state is determined by the dynamics of the vectors $\vec{r}_m$, each one 
of which lives in its own Bloch sphere. 
To obtain equations for $\vec{r}_m$ we assume that the 
Hamiltonian $h_m$ is proportional to a Pauli matrix (which induces a rotation 
of $\vec{r}_m$ on the Bloch sphere). Thus, if $h_m=\Omega\sigma_k/2$ we find 
\begin{equation}
\dot{\vec{r}}_{m}=\vec{\Omega}\times\vec{r}_m
-(\gamma(1-\delta_{m0})+N\gamma') \vec{r}_{m} +
\sum_{a=1}^N\Lambda_{a} \left(\gamma \delta_{m0} \vec{r}_{a} + \gamma' 
\vec{r}_{m+a} \right).\label{rm}
\end{equation}
Here, $\vec{\Omega}=\Omega {\bf k}$ and the matrix elements of 
the $3\times 3$ matrices $\Lambda_a$ are:
\begin{equation}
\left(\Lambda_{a}\right)_{ij}={1\over 2}\delta_{ij}
Tr(\hat\sigma_j U_a)^2.\label{etaaj}
\end{equation}
It is worth noting that, as for stabilizer codes the operators $U_a$ are Pauli 
matrices, the only possible values of the diagonal 
elements $\Lambda_{a,ii}$ are $\pm 1$.
The equations for $\vec{r}_m$ can be analytically solved in some simple cases. 
Here, we consider only the code protecting $k=1$ qubit against phase errors 
affecting  
any one of the $n=3$ carriers. In this case we can use the results 
given in the 
previous section to show that the matrices $\Lambda_{1}$ and $\Lambda_{2}$ are 
equal to the identity while $\Lambda_{3}=diag(-1,-1,+1)$. Using this, we can
obtain the equation for the polarization vector $\vec r_0$ 
(for convenience we avoid writing the subscript $0$ and 
denote $\vec{r}=\vec{r_0}$).
It turns out that $\vec{r}$ is coupled through (\ref{rm}) with two other 
vectors:
$\vec{r}_I=\vec{r}_1+\vec{r}_2$ (the sum of the two polarization vectors 
corresponding
to syndromes for which the recovery operator is $U=I$) and 
$\vec{r}_Z=\vec{r}_3$ (the
polarization vector corresponding to the syndrome for which 
the recovery operator
is $U=Z$). Thus, one can show that (\ref{rm}) reduces to:
\begin{eqnarray}
\dot{\vec{r}}&=&\vec{\Omega}\times{\vec{r}}-3\gamma' \vec{r}+(\gamma+\gamma') 
\left(\vec{r}_I+\vec{r}_Z\right), \label{eqr}\\
\dot{\vec{r}_I}&=&\vec{\Omega}\times{\vec{r}_I}-(\gamma+3\gamma' )
\vec{r}_I+\gamma'\left(2\vec{r}+2\vec{r}_Z+\Lambda_3\vec{r}_I\right),
\label{eqrI}\\
\dot{\vec{r}_Z}&=&\vec{\Omega}\times{\vec{r}_Z}-(\gamma+3\gamma')
\vec{r}_Z+\gamma'\left(\vec{r}_I+\Lambda_3\vec{r}\right). \label{eqrZ}
\end{eqnarray}
where $\vec{\Omega}=(\Omega,0,0)$. Details of the solution of this system will 
be analyzed elsewhere (Paz \& Zurek, 1997). Here we will discuss the simplest 
case where an exact solution is possible. In fact, when $\Omega=0$ 
(free evolution), the above system reduces to:
\begin{eqnarray}
\dot{\vec{r}}&=&-3\gamma' \vec{r} + (\gamma+\gamma') \vec{v}\label{eqR0}\\
\dot{\vec{v}}&=&-(\gamma+3\gamma' )\vec v+3\gamma'\vec{r} +  
2\gamma' \Lambda_3 \vec{v}\label{eqF}
\end{eqnarray}
where the vector $\vec v$ is defined as 
$\vec{v}=\sum_{a=1}^N\Lambda_{a}\vec{r}_{a}=\vec{r}_I+\vec{r}_Z$. 
Solving this system we can find 
the three components of the polarization vector $\vec{r}=(x,y,z)$. If 
the initial state belongs to the $m=0$ syndrome subspace 
(i.e., $\rho(0)=|0\rangle\langle 0|\otimes\rho_0(0)$) we find:
\begin{eqnarray}
z(t)&=& z_0 p_0(t)=
z_0\left({\gamma+\gamma'\over{\gamma+4\gamma'}}+\left(1-
{\gamma+\gamma'\over{\gamma+4\gamma'}}\right)\exp(-\lambda t)\right),
\label{z0vst}\\
y(t)&=& y_0{1\over{\lambda_--\lambda_+}}\left(
\left(3\gamma' -\lambda_+\right)
\exp\left(-\lambda_-t\right)+\left(\lambda_--3\gamma'\right)
\exp\left(-\lambda_+ t\right)\right),\label{y0vst}
\end{eqnarray}
where the decay rates are 
$\lambda_{\pm}=4\gamma'+\gamma/2\mp\sqrt{4\gamma'^2+
4\gamma\gamma'+\gamma^2/4}$ 
and $x(t)$ obeys an equation which is identical to that of $y(t)$.

Before discussing these results it is worth introducing a measure of the 
effectiveness 
of the error correcting process. For this we will use the state 
fidelity, which can 
be defined as  
\begin{equation}
F=Tr\left(\rho_{id}(t)\sum_mR_m\rho(t)R_m\right),\label{fideldef}
\end{equation} 
where $\rho(t)$ is the solution of the full master equation and 
$\rho_{id}$ is the ideal density matrix obtained by unitarily evolving
the initial state using the appropriate Hamiltonian. For the simplest
case we are considering here ($H_0=0$) the fidelity is determined by the 
scalar product 
between the initial state $\vec{r}(0)$ and the vector $(\vec{r}+\vec{v})(t)$. 
Thus, a simple calculation shows that 
$F={1\over 2}
\left(1+{1\over 2}\vec{r}(0)\left(\vec{r}+\vec{v}\right)(t)\right)$,
which, using the above results, can be written as
\begin{equation}
F={1\over 2}\left(\left(1+z_0^2\right)+\left(x_0^2+y_0^2\right)
{1\over{\lambda_- -\lambda_+}}
\left(\lambda_-\exp(-\lambda_+t)-\lambda_+\exp(-\lambda_-t)\right)\right).
\label{fidelity}
\end{equation}
A first comment on the results is that there are some states for which
the fidelity is always maximal. In fact, if the initial state is completely 
polarized 
along the $z$--axis, the code will always correct with ideal fidelity. 
The reason 
for this has nothing to do with continuous error correction but is a simple 
consequence of the fact that for such states all the errors $Z_i$ act 
trivially on the 
collective logical space $\cal{A}$. Only states which have non vanishing 
components 
on the $(x,y)$--plane are nontrivially affected by the errors we are 
considering here.   

It is useful to analyze our results in various limiting situations. For the 
simplest case with no continuous correction ($\gamma=0$), 
the fidelity is 
\begin{equation}
F={1\over 2}\left(\left(1+z_0^2\right)+\left(x_0^2+y_0^2\right)
{1\over 2}
\left(3\exp(-2\gamma't)-\exp(-6\gamma't)\right)\right). 
\label{fidelity0}
\end{equation}
This expression shows that, without continuous correction, 
the fidelity decays at a rate fixed by $\gamma'$ (the initial decay 
is quadratic because the fidelity defined in (\ref{fideldef}) includes 
a final correction event at time $t$, which is sufficient
to assure that for short times $F\approx 1-6\gamma'^2t^2$). 
The effect of continuous error correction is clearly seen when comparing
the above expression with the one arising in the limit of strong  
correction ($\gamma\gg\gamma'$) where, to first order in 
$\epsilon=\gamma'/\gamma$,
we can approximate $\lambda_-\approx\gamma(1+8\epsilon+O(\epsilon^2))$ 
and $\lambda_+\approx12\gamma\epsilon^2)$. In this case we find:
\begin{eqnarray}
z(t)&=& z_0\left(1-3\epsilon +3\epsilon\exp(-\gamma t)\right),\label{z0vst2}\\
y(t)&=& y_0\left(\left(1-3\epsilon\right)
\exp(-12\epsilon^2 \gamma t) +3\epsilon\exp(-\gamma t)\right),\label{y0vst2}\\
F&=&{1\over 2}\Big[
\left(1+z_0^2\right)+\left(x_0^2+y_0^2\right)\times\nonumber\\
&\times&\left(
\left(1+12\epsilon^2\right)\exp(-12\epsilon^2\gamma t)
-12\epsilon^2\exp(-\gamma t)\right)\Big]
\label{fvst2}
\end{eqnarray}
Two different time scales 
appear in the strong correction limit: On the one hand, 
the short time scale $t_c=1/\gamma$ (after which the 
last terms in (\ref{z0vst2}--\ref{fvst2}) are exponentially suppressed) 
is associated with the corrective jumps. Over this time scale 
the normalized 
state $\tilde{\rho}={1\over 2}\left( \hat{I}+\vec{r}
\hat{\vec{\sigma}}/p_0\right)$ 
lies on the surface of the Bloch sphere and is effectively protected 
from errors. 
On the other hand, on the much longer time $t_2=t_c/12\epsilon^2$ there is 
an overall exponential damping of the $(x,y)$--components of the Bloch vector 
which moves the normalized state $\tilde{\rho}$ towards the interior of 
the Bloch 
sphere. On this time scale the error correcting code ceases
to be efficient: fidelity decreases and purity of the quantum state is lost. 

Finally, we can briefly discuss the case where $H_0$ is nontrivial 
($\Omega\neq 0$).
For the three qubit code a fault tolerant rotation of the encoded 
state can be easily achieved by using the Hamiltonian $H_0=\Omega X_3/2$. 
Thus, as the encoding operator is such that $C X_3C^\dagger=X_3$, applying 
$H_0=\Omega X_3$ during an appropriate time interval (to make an exact 
$\pi$--pulse) we obtain an encoded bit--flip. The reason for this is 
that while the 
operator $X_3$ commutes with the generators $M_j$ (and therefore does 
not change
the syndrome) it anti--commutes with the logical pointer $L=Z_1Z_2Z_3$. 
Therefore,
applying the operator $X_3$ is tantamount to an encoded $X$ operator. With 
this Hamiltonian, the polarization vector $\vec{r}$ evolves according to 
the system (\ref{eqr}--\ref{eqrZ}) which can be numerically solved. 
As the structure of the equation is similar to the one found 
in the free case one expects the results for the fidelity to be 
essentially the same we described above. The only relevant exception is 
that the initial state with $z_0^2=1$ no longer produces ideal fidelity. 
In Figure 2 we show the time dependence fidelity for such initial state 
and display the $(y,z)$ components of the polarization vector $\vec{r}$.

We acknowledge the hospitality of the ITP at Santa Barbara
where this work was started. This research was supported in part
by the NSF grant No. PHY94--07194 and by the NSA. JPP was also supported 
by grants from UBACyT, Fundaci\'on Antorchas and Conicet (Argentina).
 
\begin{figure}
\vspace {1.0cm}
\hspace {3.0cm}
\epsfxsize=8.6cm
\epsfbox{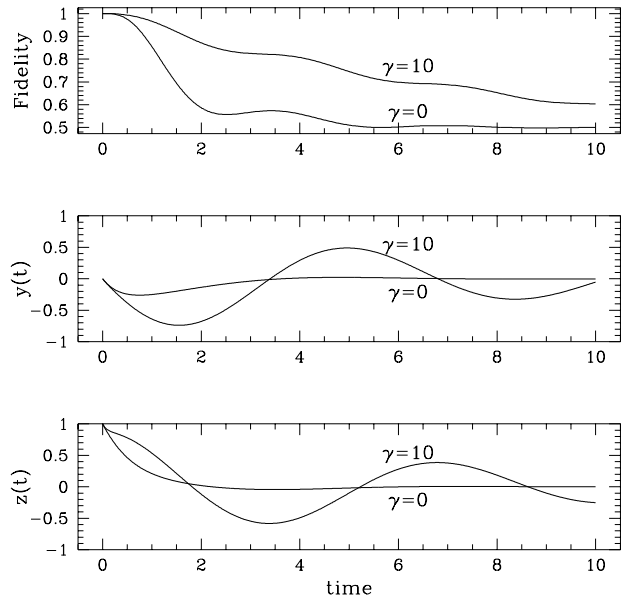}
\vspace {0.5cm}
\caption{Fidelity and non vanishing components of the polarization vector
with and without continuous error correction. The initial state is polarized
along the ${\bf z}$ axis and the Hamiltonian is $H_0=\sigma_x/2$ (the 
strength of the erroneous jumps is also set to unity).}
\end{figure}

%
%

\smallskip


%
%

\end{document}